# Photoelectrochemical water splitting with ITO/WO$_3$/BiVO$_4$/CoPi multishell nanotubes fabricated by soft-templating in vacuum

*J. Gil-Rostra [1]\*; J. Castillo-Seoane [1]; Q. Guo [2]; A. Jorge Sobrido [2]; A. R. González-Elipe [1;] A. Borrás [1]*

(1) Instituto de Ciencia de Materiales de Sevilla (CSIC-US). Avenida de Américo Vespucio, 49, 41092, Sevilla, SPAIN.

(2) School of Engineering and Materials Science, Queen Mary University of London, E1 4NS, London, UK

\*Corresponding author: jorge.gil@icmse.csic.es

ABSTRACT

A well-established procedure for the photoelectrochemical (PEC) splitting of water relies on using porous electrodes of WO$_3$ sensitized with BiVO$_4$ as a visible scavenger photoanode semiconductor. In this work, we propose an evolved photoelectrode fabricated by a soft-template approach consisting of supported multishell nanotubes (NTs). These NTs are formed by a concentric layered structure of indium tin oxide (ITO), $WO_3$, and $BiVO_4$, together with a final film of cobalt phosphate (CoPi) co-catalyst. Photoelectrode manufacturing is easily implemented at large scale and combines thermal evaporation of single crystalline organic nanowires (ONWs), magnetron sputtering (for ITO and $WO_3$), solution dripping, and electrochemical deposition processes (for $BiVO_4$ and CoPi, respectively) plus annealing under mild conditions. The obtained NT electrodes depict a large electrochemically active surface and outperform by more than one order of magnitude the efficiency of equivalent planar-layered electrodes. A thorough electrochemical analysis of the electrodes under blue and solar light illumination demonstrates the critical role of the $WO_3$/$BiVO_4$ Schottky barrier heterojunction in the control of the NT electrode efficiency and its dependence on the $BiVO_4$ outer layer thickness. Oxygen evolution reaction (OER) performance was maximized with the CoPi electrocatalyst, rendering high photocurrents under one sun illumination. The reported results demonstrate the potential of the soft-template methodology for the large area fabrication of highly efficient multishell ITO/$WO_3$/$BiVO_4$/CoPi NT electrodes, or other alternative combinations, for the photoelectrochemical splitting of water.

## INTRODUCTION

The photoelectrochemical splitting of water is a hot topic in the quest for the development of energy sustainable methods of hydrogen production. [1-3] A widely investigated procedure is the development of efficient visible light semiconductor electrodes for the oxygen evolution reaction (OER). [4-7] From the materials' composition point of view, a classical option for this purpose is the use of the wide bandgap $WO_3$ semiconductor electrode and its sensitization with an external layer of $BiVO_4$ as high performance visible light scavenger semiconductor. [8-12] Besides enabling the sensitization with blue photons of the solar spectrum, this configuration contributes to overcoming some of the problems identified by the use of $WO_3$ electrodes, namely a high photohole-photoelectron recombination rate and considerable corrosion by intermediate peroxo-like species formed during the OER. [13] Therefore, much effort has been dedicated to the tailored synthesis of layered photo-electrodes involving the stacking of these two semiconductors in shapes and configurations that maximize the effective interface area with the electrolyte. Among the rich panoply of $WO_3$ and $WO_3/BiVO_4$ high area photoelectrodes reported so far, examples exist of nanoflakes and nanorods prepared by hydrothermal methods, [14] [15] nanowires prepared by flame vapor deposition, [8] [16-19] or helix structures prepared by glancing angle evaporation, [20] the latter having reported maximum photocurrent efficiencies for this type of photoelectrodes (for a more complete appraisal of preparation methods and available structures, see the reviews in refs. [13] [21]). However, despite the high current densities already reported, most manufacturing procedures are lab scale, existing an imperious need for large-area scalable methods with portability to substrates such as carbon, nickel, titanium, or stainless steel in the form of sheets, foils, fabrics, mesh, felts, etc., which are the catalyst supports commonly used for direct electrolysis. [22] Herein we address some of these challenges by proposing the use of substrate-supported nanotubes (NTs) with a multishell nanoarchitecture including an ITO layer covered by a $WO_3/BiVO_4$ heterojunction shell bilayer. The underlying idea supporting such a concentric NT architecture with a transparent conductive layer (ITO) as the inner element is to minimize the ohmic resistance for the photoelectrons generated at the $WO_3/BiVO_4$ semiconductors providing a direct pathway toward the drain conductor layer. As it will be also shown, the arrangement in the form of nanotubes, apart from the obvious enhancement in surface area in comparison to the thin film approach, will provide a strong light scattering prompting efficient light absorption by the semiconducting shells.

In addition to the PEC evaluation of the NT electrodes, a critical point in this work has been to assess the influence of the thickness and morphology of the outer $BiVO_4$ shell layer for the oxygen evolution reaction (OER). Recently, Kafizas et al., [23] using transient absorption spectroscopy and flat $WO_3/BiVO_4$ heterojunction electrodes, have shown that photocurrent depends on $BiVO_4$ thickness, reaching a maximum for an onset value of 75 nm. Consequently, we compare herein the performance of the NT electrodes for constant ITO and $WO_3$ but variable $BiVO_4$ layer thicknesses. The selective blue light excitation of this semiconductor and the impedance spectroscopy analysis of the photoelectrodes behavior [3] [24-26] during OER have provided some clues to better understand the charge transfer mechanism and determine the optimum $BiVO_4$ layer thickness to maximize the cell photocurrent. Results have been compared with those obtained with an $ITO/WO_3/BiVO_4$ thin film (i.e. flat) electrode (denoted TF in the text) prepared by the same methodology, but presenting a much lower electrochemically active surface area. [11] [27] The optimized NT architecture has been also modified by the addition of a phosphate cobalt catalyst (CoPi) to further enhance the OER reaction efficiency. [28-32] Tested under one sun illumination, this electrode configuration rendered a photoelectrochemical current of 2.23 mA $cm^{-2}$. The obtained findings define a set of critical boundary conditions in terms of NT length, semiconductor layer thickness, and catalyst load that contribute to increasing the efficiency of nanostructured electrodes.

PHOTOANODE FABRICATION METHODOLOGY

Schematic 1 (a) presents a conceptualization of the synthetic steps required to produce the NT electrodes. The method is large area compatible and, including the incorporation of the ITO and $WO_3$ layers, can be carried out in a single vacuum deposition reactor (i.e., following a one-reactor approach). [33] The fundamentals of the procedure rely on the use of small-molecule single-crystalline nanowires (NWs) as soft-template 1D scaffolds, which can be fabricated and removed upon mild heating in vacuum. Although this is the first example showing the application of this methodology for the fabrication of a complex multifunctional architecture (i.e., $ITO/WO_3/BiVO_4$), it has a general character and has been recently applied to other oxide materials and applications. [34] [35] In steps (*i-iii*), organic nanowires (ONWs) are formed by self-assembly of pi-stacked small molecules like porphyrins, perylenes, and phthalocyanines that are vapor evaporated (EV) in vacuum on previously deposited ITO nuclei. In the present work, the NWs were made of commercially available $H_2$-phthalocyanine ($H_2Pc$). A detailed description of the mechanisms behind this process can be found elsewhere. [36] [37] A singular advantage of this protocol is its compatibility with a large variety of materials (polymers, metals, ceramics, etc.), and substrates including plastic, soda lime glass, metallic mesh, fabrics or paper, among others. In a subsequent step (*iv*) an ITO layer is deposited around the ONWs. For this step, we follow the procedure described in the reference [33] where we have reported the fabrication by magnetron sputtering (MS) of ITO nanotubes (NTs) formed on a flat ITO substrate. It is noteworthy that MS is a well-established industrially scalable technique usually employed for the fabrication of compact thin films and, more recently, also for porous structures of $WO_3$ [38] electrodes and $WO_3/BiVO_4$ photoelectrodes, [39] [40] the latter under an oblique angle configuration. The versatility of the MS procedure enables a conformal deposition of the ITO and then the $WO_3$ layers (*v*) in the same reactor (see the Materials and Methods and Supporting Information sections for details). In step (*vi*), the multishell architecture is covered with a third layer of $BiVO_4$ deposited by drop casting (DC). The annealing removal of the organic scaffold rendered substrate-supported multishell NTs (in the text denoted as NT electrodes) that have been extensively characterized and photoelectrochemically tested. Schematic 1 (b) shows a cross section representation of the sequential growth of the NT shell structure (steps (*iii*) to (*vi*)), where the resulting photoanode consists of a transparent conductive core covered by a set of concentric layers defining a semiconductor heterojunction. This configuration ensures a high surface area, the adjustment of the $WO_3$ and $BiVO_4$ semiconductor bands in line with the known electronic behavior of this heterojunction [9] [13] [16] [21] [28] [40] [41] and a minimization of the photoelectron path length to reach the ITO draining conductor layer. As a final step in the preparation procedure, a very thin layer of cobalt phosphate (CoPi) is added by photoelectrochemical activated deposition technique (c.f. Schematic 1 (b) (*vii*)).

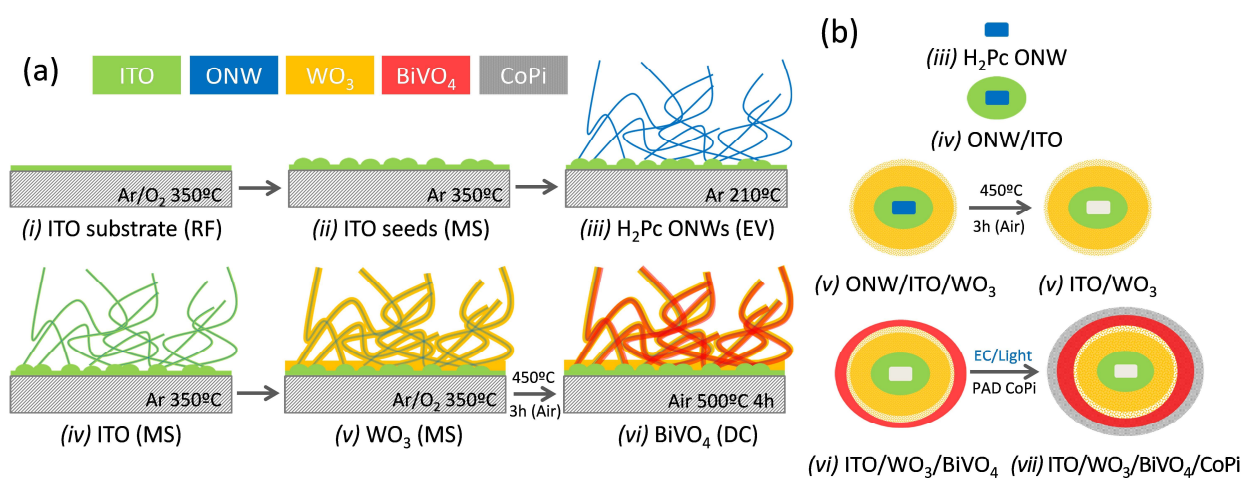

**Schematic 1.** *Conceptualization and schematic description (not to scale) of the synthesis procedure of the ITO/WO₃/BiVO₄ multishell NT photoanodes. (a) Synthesis steps of the NTs (steps i-vi) on an ITO glass substrate. (b) Schematic of the evolution of the NT cross section after the successive preparation steps (iii-vi) in (a) and after the incorporation of a Co phosphate catalyst (step vii) (see also Materials and Methods section).*

RESULTS and DISCUSSION

### Characterization and optical properties of ITO/WO₃/BiVO₄ NT electrodes

In this first part of the work, we provide information about the morphology, crystallinity, chemical composition, and optical properties of the NT samples. Figure 1 shows a series of SEM micrographs illustrating the evolution of the organic NWs and hollow NTs throughout the successive steps of the preparation protocol depicted in Schematic 1. The images correspond to normal views micrographs at different magnifications of the single organic nanowires (ONW) scaffold (Figure 1 (a)), the ONWs covered by the ITO shell (Figure 1 (b)) and the subsequent deposition of the WO₃ shell (Figure 1 (c)) and the posterior annealing treatment to completely remove the organic core formed by the ONWs. As shown below, this thermal treatment also ensures the required stoichiometry and crystalline structure of the WO₃ shell. At this stage, the image in Figure 1 (c) corresponds to a sample with a shell structure ITO/WO₃, which is labeled simply as sample NT_#0 (cf. Materials and Methods section). Figures 1 (d) and (e) show selected SEM micrographs at two magnification scales taken after deposition and annealing of the BiVO₄ semiconductor layer (samples ITO/WO₃/BiVO₄, which are labeled as NT_#1, NT_#2, and NT_#3, depending on the thickness of the BiVO₄ layer). The lower magnification image of sample NT_#2 in Figure 1 (e) clearly shows that the electrodes are formed by supported NTs with an average surface density of 5-7 NTs/μm² and lengths below 10 μm. It should be noted that the length and thickness of the NTs can be tuned by adjusting the conditions for the formation of the ONWs (for the density and length, see refs. [25] [26]), and the deposition time of the shells (for effective control of the thickness). The choice of an average length in the order of micrometers and a relatively low density of organic NWs was a compromise to maximize the conformality and thickness homogeneity of the layers prepared by MS along the length of the NTs. The comparison between Figures 1 (a) to (d) reveals that the twisted and flexible microstructure of the pristine organic templates NWs (Figure 1 (a)) transforms in a rather vertical arrangement of NTs when the ONWs become MS coated by ITO and then $WO_3$ (Figures 1(b) and (c)). This evolution has been previously reported for $TiO_2$ and ZnO NTs fabricated by the same soft template approach using plasma enhanced chemical vapor deposition and attributed to the vertical alignment of the NWs under the effect of the electrical field lines of the plasma sheath and the increase of the rigidity of the NTs after the coating by the metal oxides.[42]

Figures 1 (f) to (h) show a more detailed electron microscopy analysis (HAADF-STEM (f), bright field TEM (g) and HREM-diffraction analysis (h)) for a single nanotube retrieved from sample NT_#0 (Figure 1 (c)). This analysis reveals the hollow space left in the interior of the NTs after the annealing removal of the organic core (Schematic 1 (a) and (b) (*v*)). The empty core, as seen in Figure 1(f), appears centered and homogeneously surrounded by a radial distribution of ITO and $WO_3$. It is also apparent that the $WO_3$ NT shell was nanostructured in the form of columns perpendicular to the core axis. From an electrochemical point of view, this microstructure provides a high surface area that is expected to be beneficial for photoelectrochemical efficiency. The bright field image in Figure 1 (g) illustrates a good definition of the $ITO/WO_3$ interface, while the high-resolution electron micrograph (HREM) in Figure 1 (h) reveals the polycrystalline character of the grains and their random orientation. In this HREM image and the inset showing the electron diffraction diagram, it is possible to recognize the diffraction features and interplanar distances typical of the crystalline planes of $WO_3$. It is also apparent in the image in Figure 1 (f) that the NT width slightly varies from its top to its bottom, the latter corresponding to its anchoring to the ITO substrate. The shape of NTs is a result of the characteristics of the MS technique together with the self-shadowing effects known on the deposition of 1D scaffolds. [33] Depending on conditions, some particles deposited by physical vapor deposition may be ballistic, undergo shadowing effects at the landing position on the surface and preferentially accumulate at the tip parts of nanostructures. [43] To minimize this effect and to achieve the maximum possible homogeneous distribution of material along the whole NT length, it was required to use deposition conditions favoring the particle randomization through collisions with the plasma gas (essentially, relatively high pressure and separation target-substrate). [44]

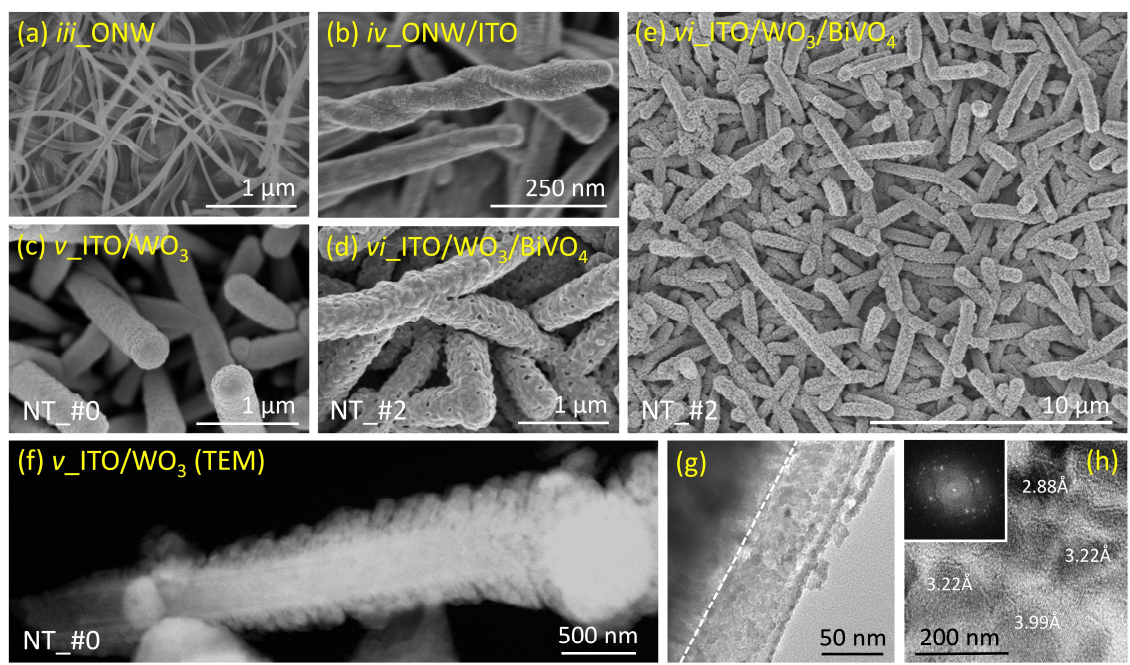

**Figure 1**. *(a)-(e) Selected normal view micrographs at various magnification scales of the hollow multishell NTs at different stages of their manufacturing process according to Schematic 1. (f) HAADF-STEM, (g) bright field TEM and (h) HREM-diffraction analysis*

*of a single NTs retrieved from sample NT_#0. Please note that labels indicate the different steps detailed in Schematic 1.*

The following step of the photoanode manufacturing process consists of the deposition of $BiVO_4$ by drop-casting (Schematic 1 (a) (*vi*)). Figure 2 (a) to (e) show characteristic SEM micrographs at low and high magnifications of the NT structures of samples NT_#1, NT_#2 and NT_#3. Figure 2 (f) presents a HAADF-STEM image of an individual NT taken from sample NT_#1. This image clearly proves that the microstructure of this layer is formed by grains covering the NT outer surface (see also Figure S1 in the Supporting Information section for more detailed cross-section views of samples NT_#1 and NT_#3 obtained after removing the domed tips of the nanotubes by FIB). As expected, the average NT thickness increases with the amount of $BiVO_4$ (i.e., with the number of drop casting processes), reaching an average thickness higher than 500 nm in sample NT_#3, corresponding to the thickest shell of the photosensitizer oxide. Micrographs in Figures 2 (a) to (c) demonstrate a progressive homogenization of the $BiVO_4$ layer from sample NT_#1 to sample NT_#3. Thus, in sample NT_#1, $BiVO_4$ forms a very thin layer (thickness in the order of a few tenths of nanometers) filling the porous structure of the $WO_3$ layer, together with small imperfectly coalesced aggregates. EDX maps in Figure S1 confirm that in sample NT_#1, $BiVO_4$ spreads relatively well onto $WO_3$, while in samples NT_#2 and NT_#3, $BiVO_4$ forms larger aggregates partially connected in a kind of granular shell (Figures 2 (c) to (d)).

Although an accurate estimation of the NT thickness is hampered by the inherent random distribution of NT sizes, a statistical analysis of the SEM images taken at medium magnifications for a selected set of representative NTs (see data gathered in Table S1 of the Supporting Information) gives the following average values of thickness: $438 \pm 64$ nm (sample NT_#0), $483 \pm 35$ nm (sample NT_#1), $568 \pm 78$ nm (sample NT_#2) and 582 ± 48 nm (sample NT_#3). These measurements render approximate thickness values for the BiVO$_4$ layer of 30 nm, 65 nm and 85 nm for samples NT_#1, NT_#2 and NT_#3, respectively. For comparison, the thickness of the layers in sample TF derived from the cross-section micrograph in Figure S2 present mean values of 550 nm and 370 nm for the WO$_3$ and BiVO$_4$ layers, respectively.

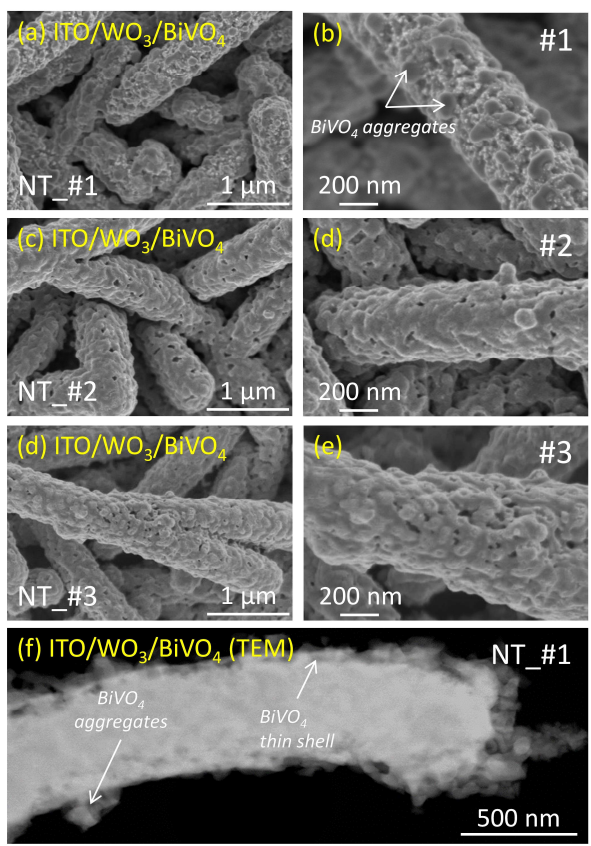

**Figure 2**. *SEM micrographs (a)-(d) at two different magnifications (1μm and 200 nm scale) of the multishell NTs, corresponding to increasing amounts of BiVO$_4$ for samples NT_#1, NT_#2 and NT_#3, as labeled. d) HAADF-STEM image of a single NT taken from sample NT_#1.*

The crystal structure of WO$_3$ and BiVO$_4$ semiconductors is of paramount importance for the performance of WO$_3$/BiVO$_4$ heterojunction photocells. [8-12 16 21 40 41] An analysis of the crystallinity of the NT electrodes by XRD rendered the diagrams in Figure 3 (a) for samples NT_#0, NT_#1, NT_#2 and NT_#3, where they develop the typical diagram peaks of well crystallized ITO (cubic structure), WO₃ (monoclinic structure), and BiVO₃ (monoclinic structure). [45-47]

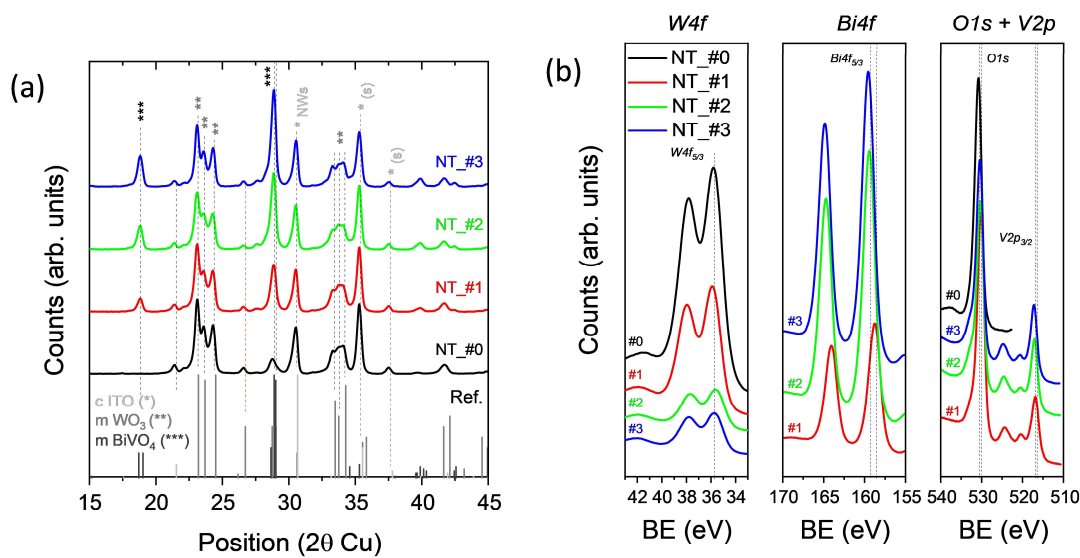

**Figure 3.** a) *XRD diagrams for samples NT_#0 to NT_#3. Peaks attributed to the cubic structure of In₂O₃/SnO₂ (\*) [45] the monoclinic structure of WO₃ (\*\*) [46] and the monoclinic structure of BiVO₄ (\*\*\*) [47] can be identified in the diagrams. b) W4f, Bi4f and O1s/V2p photoelectron spectra of samples NT_#0, NT_#1, NT_#2 and NT_#3.*

The effect of the deposition of BiVO₄ by drop-casting and the chemical changes that occur at the surface of NTs were investigated by XPS and EDX. The comparison of general spectra reported in Figure S3 shows a progressive decrease of the W4f/Bi4f intensity ratio observable when passing from sample NT_#0 to samples NT_#1, and NT_#2, with little observed differences in this parameter for sample NT_#3. This evolution confirms the previous assessment by SEM observation in the sense that the coverage of the sensitizing BiVO₄ semiconductor onto the WO₃ shell is rather conformal in samples NT_#1 and NT_#2, while in sample NT_#3 there is a certain agglomeration of BiVO₄ in the form of relatively bigger clusters. The fact that the W4f signal is still visible in samples NT_#1, NT_#2 and even NT_#3 also indicates that some zones of WO₃ are not completely covered by $BiVO_4$. Meanwhile, the series of zone spectra gathered in Figure 3 (b), besides supporting this assessment about coverage based on the general spectra, confirm the $W^{6+}$, $Bi^{3+}$ and $V^{5+}$ oxidation states of these elements (B.E.s W4f (36 eV), Bi4f (159 eV) and V2p (517 eV)).

In parallel to the changes outlined above on morphology and surface chemistry, samples underwent color changes after each processing step. In particular, a white-yellow transformation was observed between steps (*v*) and (*vi*) that must respond to the different diffuse reflectance spectra of samples NT_#0 (with only ITO and $WO_3$ layers after annealing) and NT_#1 to NT_#3 after the incorporation of $BiVO_4$ (see Supporting Information, Figure S4). Most remarkable in these spectra is the shift in the absorption edge from sample NT_#0 (approximately 2.7 eV) to samples NT_#1-#3 (approximately 2.4 eV for the three samples). This shift clearly proves that the incorporation of $BiVO_4$ produces the yellow coloration of the samples due to the capacity of this semiconductor to absorb visible light in the wavelength interval comprised between, approximately, 400 and 450 nm. Most experiments described from now on were carried out by illuminating the cell with photons within this energy window and therefore selectively inducing the photon excitation of the $BiVO_4$ semiconductor.

As shown in the next section, NT electrode samples NT_#1-#3 depict photoelectrochemical efficiencies that are orders of magnitude higher than that of the reference sample TF. This enhancement in photocurrent may depend on the increase in surface area provided by the NT structure of samples NT_#1-#3, although a certain enhancement in light scattering within these samples cannot be discarded due to the random arrangement of NTs [3] (this feature has been recently demonstrated by us for ITO NT structures). [33] The importance of the increase in surface area in samples NT_#0-#3 as compared with sample TF was proved by determining the effective electrochemically active surface area (ECSA) using the double-layer capacitance method. [48-52] This analysis entails determining the slope of the current density plots vs. scan rate (see Supporting Information Figure S5). The obtained capacitance values revealed that the multishell NT electrodes depict a doble layer capacitance almost two orders of magnitude higher than that of the reference TF. Also interesting for the evaluation of photoelectrochemical response is that the incorporation of $BiVO_4$ does not significantly alter the capacitance values of NT samples. In other words, the samples NT_#0 to NT_#3 exhibit rather similar ECSA values.

***Photoelectrochemical activity of thin film, multishell NTs and CoPi modified electrodes.***

Voltammetric and chronoamperometric tests were carried out on the NT and NT-CoPi electrodes to assess their photoelectrochemical activity in comparison with the TF samples. Most tests were carried out upon illumination provided by a 6500K LED, although experiments were also performed under illumination with a solar simulator in order to determine quantitatively the response of the most efficient NT electrodes.

The photoelectrochemical performance of NTs and TF electrodes was firstly assessed by linear sweep voltammetry (LSV) analysis. Figure 4 (a) shows the LSV curves measured under 6500K LED illumination with a light flux power of 100 mW cm$^{-2}$ for samples NT_#0-#3 and TF_#2. Clearly, samples NT_#0 and TF_#2 present a much lower photoelectrochemical efficiency than samples NT_#1 to NT_#3. We relate this difference with the absence of the sensitizing $BiVO_4$ semiconductor in sample #0 and the significantly lower ECSA and light dispersion capacity of the thin film configuration. From this plot and the comparison of efficiencies for samples #1 - #3 it is also apparent that efficiency is maximum for sample NT_#2. A more complete LSV analysis of the photoelectrochemical response for the multishell NT electrodes as a function of the light flux is reported as Supporting Information Figure S6). This complementary analysis showed that no saturation of current density exists even for light intensities as high as 300 mW cm$^{-2}$ and that, even at very high photon fluxes, the photoelectrochemical currents determined for sample TF_#2 and sample NT_#0, were comparatively very small.

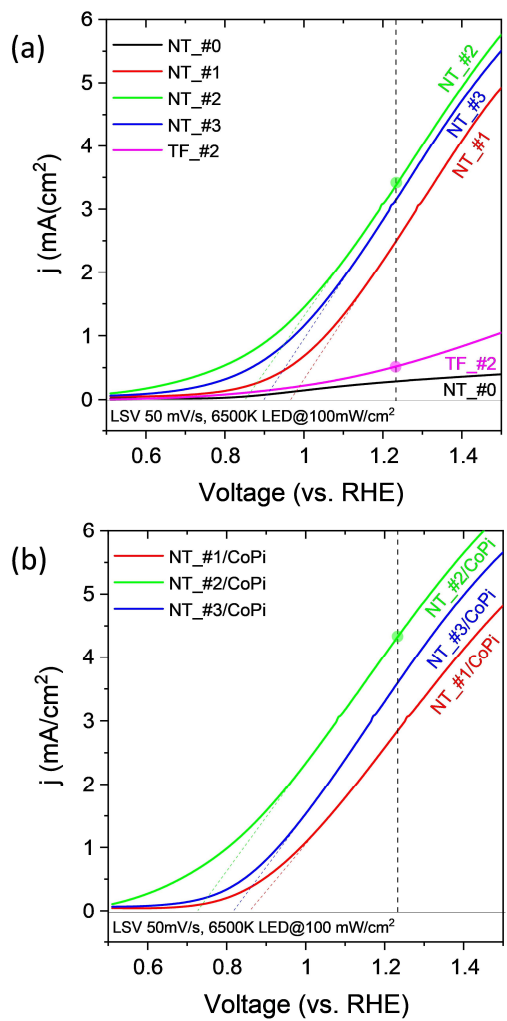

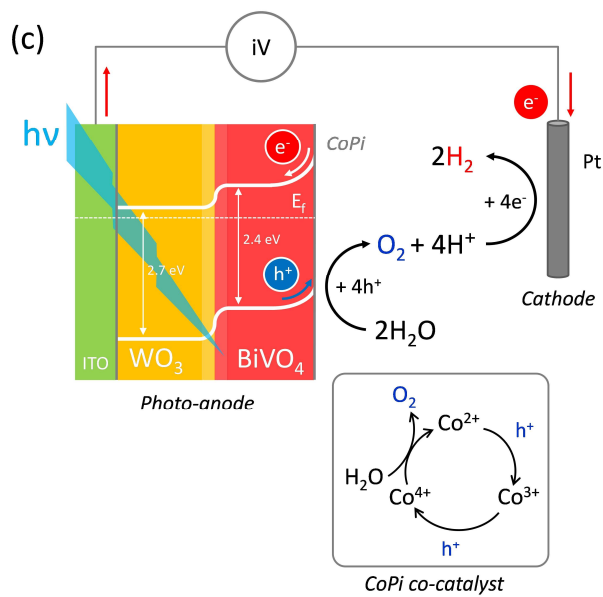

**Figure 4.** *LSV analysis of NT electrodes under light excitation. (a) LSV diagrams under illumination with blue light for multishell NTs (i.e.NT_#0, NT_#1-#3) and TF_#2 electrodes. (b) LSV diagrams under illumination with blue light for NT_#1/CoPi, NT_#2/CoPi and NT_#3/CoPi electrodes. (c) Band diagram describing the fate of photoelectron and photohole upon light absorption by the BiVO₄ semiconductor. Insert: schematic description of the catalytic mechanism of the Co-Pi catalyst to activate the OER.*

The observed photocurrent of samples NT_#1-#3 responds to the visible absorption activity of the $BiVO_4$ layer. According to current models of photoexcitation of sensitizing semiconductors, [8-12] photon absorption by the $BiVO_4$ produces electron-hole pairs. As shown in the scheme in Figure 4 (c), the Schottky band bending of the $BiVO_4$ semiconductor close to the water-photocatalyst interface promotes the diffusion of the photoholes to the surface where they react with water to yield a series of very reactive intermediate species (e.g., OH*, OOH*, O*), [53] [54] ending up in the generation of $O_2$. Meanwhile, the photo-electrons that have been promoted to the conduction band diffuse to the valence band of $WO_3$ and, then, through the ITO conductive layer, to the external circuit and eventually to the cathode to reduce water into hydrogen. The high efficiency found for this process for the NT electrodes supports a good matching at the heterojunction between the valence and conduction bands of the two semiconductors (see Figure 4 (c)) and the fact that, even for the proposed NT nanoarchitecture, photoelectrons only need to diffuse through the $WO_3$ shell thickness to find the ITO draining layer. Interestingly, equivalent experiments with green and red radiations (centered at 520 nm and 635 nm, respectively) did not produce any significant photoelectrochemical reaction, in agreement with the negligible photon absorption coefficient of $BiVO_4$ at these wavelengths (data not shown).

The efficiency for the OER further increased when the nanotubes are decorated with the CoPi catalyst. The LSV plots recorded in Figure 4 (b) for samples NT_#1/CoPi-#3/CoPi reveal an increase in current density for these three samples with respect to the photoelectrochemical response of samples without the OER catalyst. According to the literature, the OER catalytic activity of the cobalt involves the formation of intermediate species of $Co^{3+}$ and $Co^{4+}$, the latter in the form of CoO(OH) at high positive voltages (V> 1.23 V) (see the scheme in Figure 4 (d)). [55] Interestingly, the plots in this figure confirm the order for the efficiencies found for the NT electrodes without catalyst, i.e., #2/CoPi > #3/CoPi>#1/CoPi.

The same order in efficiency (i.e. samples #2> #3 > #1) found by the LSV analysis was determined by long term chronoamperometry analysis (CA). Figure 5 (a) shows the CA curves recorded as a function of time for an applied voltage of 1.23 V vs. reversible hydrogen electrode (RHE). Results correspond to the NT electrodes with the CoPi co-catalyst, although curves showing the same trend were also obtained without co-catalyst. According to these curves, photocurrent intensity depicts a first increase in photocurrent for the first 30 min, followed by a smooth decrease during the following 3 hours. Similar behavior has been reported [56] and attributed to the effective and reversible formation of $Co^{3+}$ and $Co^{4+}$ species during the first activation period followed by a subsequent stabilization process.

A summary of average photocurrent values, determined by LSV and CA experiments for the NT electrodes with and without CoPi is gathered in Figure S7 corroborating that samples NT_ #2 and NT_#2/CoPi depict the maximum efficiency, in agreement with the results in Figures 4 and 5. Interestingly, when the used samples were taken out of the liquid medium, dried, and stored for three months to then perform a new experiment, the evolution of the potential with time depicted in Figure 5 (a) was completely reproducible. This reproducibility proves the stability and robustness of the nanowire electrodes and their long lasting stability.

The light induced character of the electrolysis and its long-term stability were confirmed by complementary CA tests where 6500K LED light was systematically chopped at small time intervals. Figure 5 (b) illustrates the type of behavior found for an experiment carried out with the sample NT_#2/CoPi. This figure shows that current is illumination-dependent and that intensity recovers without practically any degradation after multiple chopping processes.

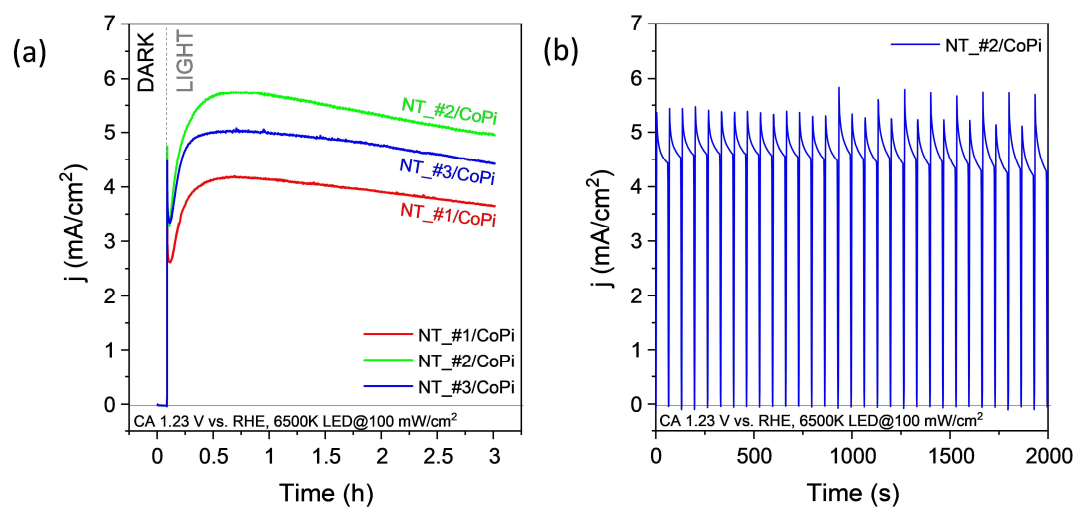

***Figure 5.*** *Chronoamperometry tests of NT electrodes with CoPi catalysts. (a) CA diagrams at 1.23 V (vs. RHE) recorded for a long time for samples NT_#1/CoPi, NT_#2/CoPi and NT_#3CoPi. (b) CA curves chopping the light illumination approx. every 60 s. Plot corresponds to sample NT_#2/CoPi.*

A detailed analysis of the CA plots in Figure 5 (b) can be used to determine transient times for the photo-hole migration through the $BiVO_4$ semiconductor. According to the works in refs. [57] [58], the analysis of the initial transient decay up to stabilization of the current curve after chopping illumination (see Figure S8) can provide an estimate of the average diffusion time of photo-holes in their path to reach the surface. The diffusion time ($\tau$) can be estimated according to the following expressions: [57] [58]

$$lnD = -\frac{1}{\tau}t \quad \text{(Eq. 1)}$$

$$D = \frac{j(t)-j_f}{j_f-j_i} \quad \text{(Eq. 2)}$$

Where D can be determined experimentally as indicated in Figure S8. Calculated values of the diffusion time ($\tau$) amounted to 9.5 µs for sample NT_#0 (note however, the small current generated with this electrode, c.f, Figure 5 (a) and Figure 6 (a)) and 12.8 µs for the three multishell NT_#1/CoPi-#3/CoPi electrodes. These values are in good agreement with the previously reported for other nanostructured electrodes. [21 57 58] The similar photo-hole diffusion values found for the three multishell NT electrodes suggest that the observed differences in efficiency must involve additional factors, such as reaction kinetics at the electrode interface and other charge transfer phenomena that will be discussed in the next section.

In order to compare the efficiency of the NT electrodes with that of other electrode morphologies reported in the literature, [8-12] LSV diagrams were also recorded under illumination with a solar simulator lamp for electrodes NT_#2 and NT_#2/CoPi and, for comparison, also for electrode NT_#0. Figure 6 (a) shows that electrode NT_#2 is much more effective than electrode NT_#0, although some intensity could be recorded for this latter sample illuminated with the complete solar spectrum. Figure 6 (a) also shows a net increase in efficiency when incorporating the CoPi catalyst (sample NT_#2/CoPi). For this electrode and one sun illumination, the photocurrent obtained amounted to 2.23 mA at 1.23 V vs RHE. It is also noteworthy that the onset potential for the OER has a remarkably low value of 0.63 V for this electrode. On the other hand, the IPCE curves in Figure 6 (b) confirm that samples NT_#2 and NT_#2/CoPi are effectively sensitized with photons with λ<500 nm, while sample NT_#0 presents photoresponse only for photons with λ<450 nm.

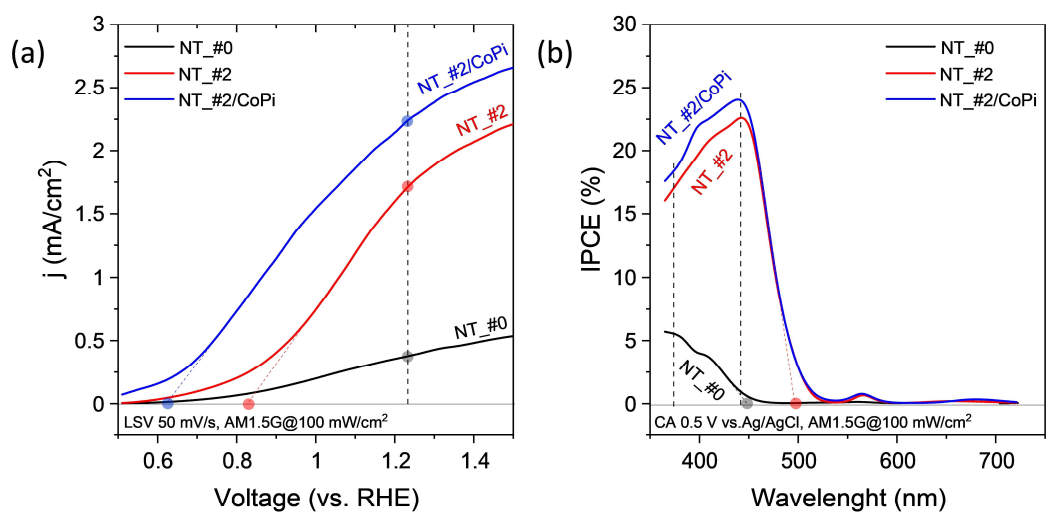

**Figure 6.** *Electrode sensitization with a solar simulator. (a) LSV plots of samples NT_#0, NT_#2, and NT_#2/CoPi. (b) IPCE curves for these same electrode samples.*

### Electrode morphology and charge transfer processes at the NT electrode-electrolyte interface

To account for the nanostructure and morphology factors of NT electrodes that affect the photoelectrochemical cell efficiency and their superior performance to that of flat TF electrodes, we have characterized the electrical behavior of the cell by electrochemical impedance spectroscopy (EIS) and assessed by means of a Tafel analysis the kinetics of the charge transfer process at the NT and TF electrode interfaces, with and without CoPi co-catalyst. As a result, we have gained a comprehensive understanding of the main factors contributing to the high efficiency of the NT electrodes and some clues to account for the higher efficiency of samples NT_#2 and NT_#2/CoPi in terms of the equivalent thickness of BiVO₄ layer in this sample.

Nyquist plots obtained by impedance spectroscopy for the different anodes are reported in Figure 7 (a). The plots can be fitted by assigning specific values to the electrical elements of the equivalent circuit included as an inset in this figure. Such a circuit type has been previously proposed to account for the charge transfer behavior for nanostructured $WO_3/BiVO_4$ heterojunction electrodes. [24-26]

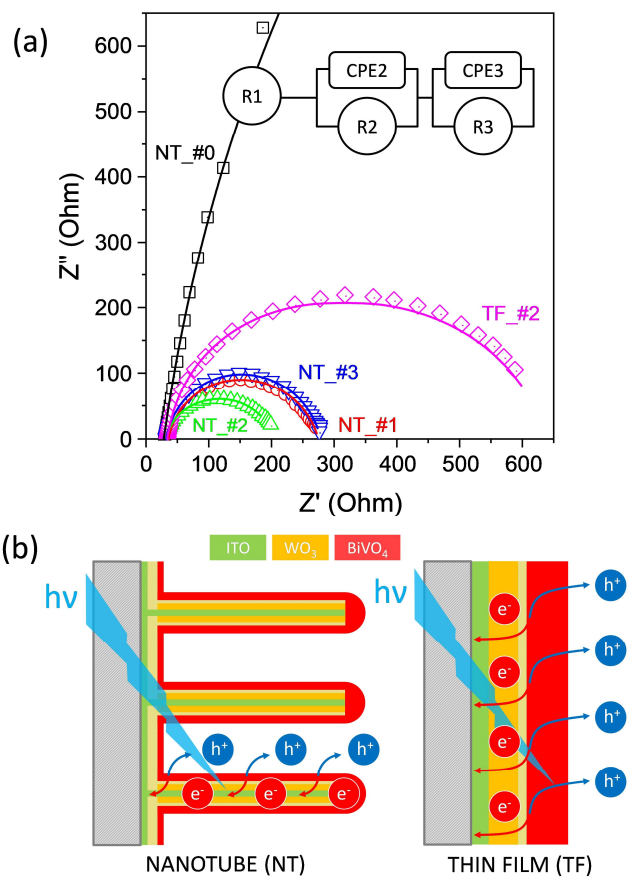

**Figure 7**. *(a) Nyquist plots for samples NT_#0-#3 and TF_#2 recorded under 6500K LED illumination with an intensity of 100 mW/cm² and an applied voltage of 1.23 V vs. RHE. Inset: scheme of the equivalent circuit. (b) Scheme of the nanostructure of NT and TF electrodes and the involved photoelectrochemical process highlighting the operative advantages of the ITO/WO₃/BiVO₄ nanotube nanostructure.*

This equivalent circuit incorporates three different elements: i) element 1 characterized by a resistance R1 that is attributed to the resistance of the electrolyte solution, wires, clips and ITO substrate layers; ii) element 2 consisting of a resistance R2 and a constant phase element CPE2 that can be attributed to the contribution of the interfaces between the various oxides integrated in the multishell structure of the NTs and TF samples; iii) element 3 including a resistance R3 and a constant phase element CPE3 associated to the charge transfer at the $BiVO_4$/electrolyte interface upon light illumination. Phase constant elements instead of pure capacitance elements are used in the scheme because non-linearity effects often appear at heterogeneous and nanostructured surfaces and interfaces. [24 59 60] The impedance associated with the CPE can be formulated according to:

$$Z_{CPE} = \frac{1}{Q\,(i\omega)^n} \qquad \text{(Eq. 3)}$$

were $Q$ is the CPE parameter in F·s, $n$ is a number between 1 and 0, and $\omega$ is the angular frequency. The set of Nyquist plots reported in Figure 7 have been fitted using the parameters summarized in Table 1 (for the quality of the fitting results see continuous lines in Figure 7 (a)). These parameters encompass calculated values of resistance $R$, $Q$ (directly related to the non-ideal capacitance characteristic of the said element), and an ideality factor ($n$) for each constant phase element.

**Table 1.** *Fitting parameters reproducing the shape of the Nyquist plots recorded at 1.23 V vs. RHE under LED 6500K illumination (100 mW/cm$^2$)*

|  |  | NT_#0 | NT_#1 | NT_#2 | NT_#3 | TF_#2 |
|---|---|---|---|---|---|---|
|  | **R1** (*Ohm*) | 28.2 | 35.3 | 31.75 | 32.3 | 33.6 |
|  | **R2** (*Ohm*) | 2011.4 | 204.9 | 150.4 | 147.7 | 429.5 |
|  | **R3** (*Ohm*) | 2530.5 | 34.7 | 22.5 | 98.3 | 175.4 |
| **CPE2** | Q2 (*F·s*/10$^{-4}$) | 2.4 | 3.0 | 4.7 | 5.1 | 5.0 |
|  | n2 | 1.00 | 0.82 | 0.79 | 0.85 | 0.81 |
| **CPE3** | Q3 (*F·s*/10$^{-4}$) | 2.3 | 3.8 | 3.9 | 3.2 | 2.0 |
|  | n3 | 0.86 | 1.01 | 1.01 | 0.93 | 0.99 |

R1 parameters are rather similar for the different studied samples, confirming the attribution of this resistance to the electrolyte and other ohmic elements present in the external circuit connecting photoanode and cathode. [24] On the other hand, R2 and R3 and their sum for electrode NT_#0 are more than one order of magnitude higher than for the other NTs and TF electrodes proving that the incorporation of a $BiVO_4$ semiconducting outer layer is responsible for the photogeneration of charge upon light excitation. CPE2 and CPE3 are also higher for these electrodes than for electrode NT_#0. However, the most remarkable from this fitting analysis is that R2+R3 values for the NT electrodes are much smaller than in the TF electrodes, supporting the dependence of overall photoelectrochemical activity on the effective ECSA available in each case. As already pointed out in previous sections, an additional factor contributing to reducing the value of R2+R3 for the NT electrodes resides in that the inner ITO layer in the nanotube structure decreases the resistance to the path and collection of photo-generated electrons in the $BiVO_4$ photoactive component. In summary, as schematized in Figure 7 (b), the core-shell $ITO/WO_3/BiVO_4$ nanostructure of the NT electrodes offers simultaneously a high electrochemical surface for the OER and minimum electrical resistance losses due to its particular core-shell configuration.

A Tafel analysis, based on the Butler-Volmer equation, [52] has been used to estimate the kinetics of charge transfer processes at the electrode-electrolyte interface. Figure 8 (a) shows a series of Tafel plots determined for the NT, NT-CoPi and TF electrodes. The slope of these plots directly relates to the rate of the charge transfer processes at the interface. According to this figure, slopes of electrodes NT_#0 to NT_#3 are quite similar but bigger than that determined for electrodes NT_#1/CoPi to NT_#3/CoPi, clearly indicating that the addition of the CoPi catalyst makes more effective the charge transfer process at the interface. We assume that for these samples a majority of photoholes arriving at the interface end up in the formation of $O_2$ molecules. The Tafel slope for the TF sample was higher than the NT electrodes, supporting a lower reactivity on this structure of the photoholes to generate $O_2$ molecules. This result is consistent with the lower photocurrent efficiency found for the TF configuration (see Figure 4), which is also clearly related to the significantly low ECSA values shown by the TF_#2 sample (see Figure S5). As spected, and for a given $BiVO_4$ semiconductor load (i.e. #2), a flat layered structure presents less active surface area to efficiently transfer the generated photoholes compared to an extensive and well distributed multishel NT structure. Interestingly, despite the similar Tafel slopes for the NT and NT-CoPi electrodes, the log j values at a given potential follow the order #2>#3>#1 (and also for their equivalent electrodes with CoPi). Taking into account the similar photohole diffusion times determined according to equation (1) in the previous section (see Figure S8) and the similar interface reaction rates of photoholes suggested by the Tafel slope analysis (c.f., Figure 8), the maximum efficiency found for sample NT_#2 (and NT_#2/CoPi) suggests that the number of photoholes arriving at the interface reaction sites is higher for this particular electrode. We propose that the observed variation of efficiency with the $BiVO_4$ load can be related to the different dispersion degree of $BiVO_4$ phase and the relation between the length of the depletion layer (i.e, length of the Schottky barrier defined by the band bending zone) and the equivalent thickness of the $BiVO_4$ sensitization layer/aggregates in samples NT_#1 to NT_#3. According to the schemes in Figure 8 (b) the semiconductor sensitizing layer/aggregate would reach a critical thickness in sample NT_#2 enabling the complete development of the Schottky barrier through the semiconductor layer. Under these conditions light absorption and separation of photogenerated electrons and holes would be quite effective since all photogenerated holes would experience the electrical field of the band curvature, thus diminishing the probability of photoelectron-photohole recombination. A similar effect was reported by us to account for the observed enhancement in photocatalytic activity of rutile-anatase bilayer system. [61] Interestingly, the average thickness determined for this layer by the statistical morphological analysis of NTs in sample NT_#2 is about 65-85 nm, a similar range value to that found by Selim et al. [23] (<125 nm) as the onset for a stable photoresponse in a series of $WO_3/BiVO_4$ flat layer photoelectrodes where the $BiVO_4$ thickness varied from 75 to 350 nm. According to the scheme in Figure 8 (b), the equivalent thickness of the $BiVO_4$ semiconductor in sample NT_#1 (ca. 35 nm, see Figure 2 and Table S1) would be smaller than the charge space length, as estimated for sample NT_#2, making the photoelectron/photohole separation less effective. Meanwhile, in sample NT_#3, $BiVO_4$ form bigger aggregates approaching a 3D particle morphology (see discussion in Section 2.1 ) with particle sizes larger than the Schottky barrier length. For this morphology the recombination probability of photogenerated electrons and holes would be higher in the zone where bands are flat, leading to a decrease in the number of photoholes effectively arriving at the surface.

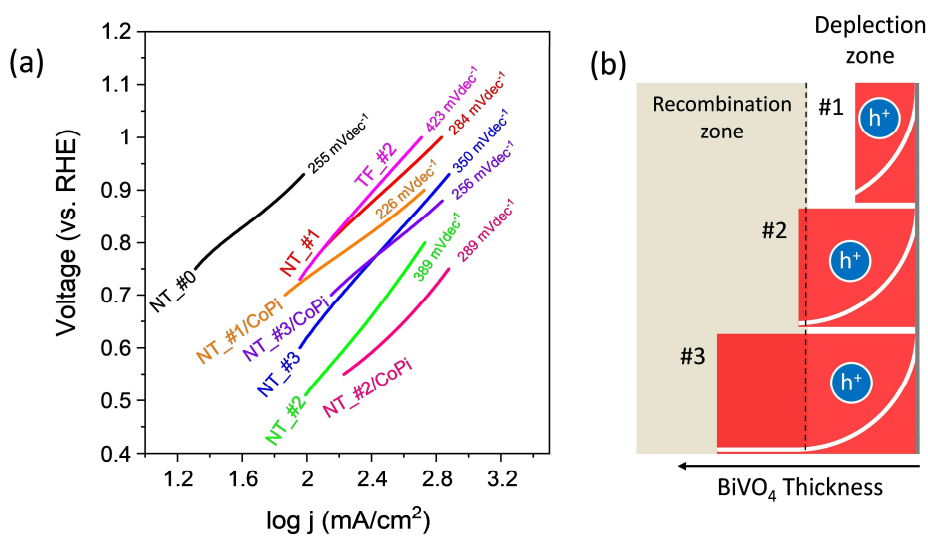

**Figure 8.** *Control of the charge transfer process at the electrode/electrolyte interface. (a) Tafel plots for NT, NT/CoPi and TF electrodes. (b) Band bending schemes for samples*

*#1, #2 and #3 taking into account the estimated thickness of BiVO₄ layer/aggregates and how it affects the Schottky barrier length.*

## MATERIALS AND METHODS

### *Synthesis of multishell NT photoelectrodes*

The multishell nanotube electrodes were manufactured following the steps described in Schematic 1. Deposition conditions of ITO and $WO_3$ layers by MS and the processing conditions of $BiVO_4$ coatings and CoPi layers by solution dripping or electrochemical methods are detailed in Supporting Information section S1. We used a commercial glass plate covered by ITO layers as substrates (supplied by Ossila, 400 nm thickness, 15 Ω/□). ITO substrates were ultrasonically bath cleaned following a standard acetone, ethanol, deionized (DI) water solvents sequence. Once in the vacuum chamber the substrates underwent a cleaning and surface activation process by a mild radio frequency (RF) plasma treatment in an $O_2$/Ar atmosphere (step (*i*) in Schematic 1 (a)). Small ITO nuclei were additionally deposited by MS for very short times (step (*ii*)). These ITO nuclei served as nucleation seeds to facilitate the crystalline formation of the organic phthalocyanine nanowires (step (*iii*)). Details of the experimental procedure for the formation of the ONW templates by thermal evaporation can be found in previous publications [33] [36] [37] (see also Supporting Information section S1). Then, ITO was deposited by MS under operating conditions leading to a conformal layer completely covering the organic nanostructured templates (step (*iv*)). This first layer deposition of ITO was followed by the MS deposition of $WO_3$ (step (*v*)). The steps described so far have been carried out following a one-reactor approach, i.e. using solely a vacuum reactor chamber including the thermal evaporation and magnetron sputtering sources. This protocol allows us for a rapid deposition avoiding transference of samples and the exposure of the successively formed surfaces to ambient conditions, appealing characteristics when aiming for large-scale deposition and finely controlled interfaces. After vacuum deposition, ITO/$WO_3$ structures were annealed in air at 450 ºC for 3 h to achieve the complete removal of the organic scaffold at the core of the NTs and to induce the crystallization of the $WO_3$ semiconductor oxide (step (*v*)).

Finally, $BiVO_4$ was incorporated by sequentially dripping on a delimited substrate area of 1.1 cm$^2$ different volumes (i.e. 20, 40, 60 µl) of a solution of $BiNO_3x5H_2O$ (50 mM), $VO(acac)_2$ (46.5 mM) and an acetic acid/ethyl acetate (9.5/0.5) mixture as a solvent. After the complete removal of the solvent (at room temperature) the samples were annealed in air at 500 ºC for 4 h to induce the complete crystallization of $BiVO_4$ and promote the $WO_3$/$BiVO_4$ heterojunction formation (see step (*vi*)). The final result is a hollow multishell NT structure as indicated in Schematic 1 (b) step (*vi*) where the $BiVO_4$ shell semiconductor covers externally the $WO_3$ (and the ITO) concentric shells. As mentioned, the dripping process was repeated up to three times to check the influence of the amount of $BiVO_4$ on the photo-efficiency of the electrodes. The set of fabricated $WO_3$/$BiVO_4$ photocatalyst samples will be named as NT_#0 for the ITO/$WO_3$ single semiconductor system, and NT_#1, NT_#2, and NT_#3 for the ITO/$WO_3$/$BiVO_4$ heterojunction system fabricated by adding $BiVO_4$ precursor solution volumes of 20 µl, 40 µl and 60 µl, respectively. A reference flat thin film sample (TF) was prepared in the form of stacked thin films applying the already described layer deposition steps (i.e., dripping 40 ml of the $BiVO_4$ precursor solution) directly onto a flat ITO substrate with no organic scaffold nanostructure (sample TF_#2).

The cobalt phosphate (CoPi) co-catalyst was deposited through a light stimulated electrochemical deposition method.[32] For that, the electrodes were back illuminated with the light provided by a solar simulator (100 mW/cm2) in a solution of 0.15 mM $Co(NO_3)_2x6H_2O$ in a 0.1 M potassium phosphate buffer (at pH=7.0) applying a constant voltage of 0.4 V (vs. Ag/AgCl (3M KCl) reference electrode) for 600 s. Following the proposed notation, the CoPi set of samples has been labeled as NT_#X/CoPi (X: 1, 2 and 3).

### *Characterization of multishell NT photoelectrodes*

The microstructure of the NTs was characterized by scanning electron microscopy (SEM) and transmission electron microscopy (TEM) together with EDX analysis, the latter to ascertain the layered structure of the multishell NTs. The Hitachi S4800 SEM working at 2kV was utilized for the normal view characterization of the electrodes). Additional morphological and chemical characterization was obtained at the Helios Nanolab 650 FIB-SEM from FEI equipped with an EDX detector from Oxford (Microscopy Service from the University of Malaga). A Tecnai F30 TEM instrument working a 300 kV was utilized for the HAADF-STEM imaging, SAED, and HREM analysis.

Crystalline structure of the electrodes was ascertained by X-ray diffraction in a Panalytical X'PERT PRO model operating in the θ - 2θ configuration and using the Cu Kα (1.5418 Å) radiation as an excitation source.

X-ray photoelectron spectroscopy (XPS) analysis of the outer layers of the electrodes to determine the $WO_3/BiVO_4$ coverage ratio and to confirm the metal chemical state of the oxide semiconductors was carried out in a PHOIBOS 100 hemispheric multichannel analyzer from SPECS using the Al Kα radiation as an excitation source. General and peak spectra were acquired with pass energies of 50 and 30 eV, respectively. Spectra were calibrated positioning the C1s peak at 284.5 eV.

UV-vis-NIR absorption spectroscopy analysis was carried out recording the diffuse reflectance spectra of the electrodes in a PerkinElmer LAMBDA750S spectrometer equipped with a 60 mm diameter integrating sphere.

**Electrochemical and photo-electrochemical analysis and performance**

Photoelectrochemical analysis of the multishell NTs and thin film TF reference electrodes was carried out in a three electrodes cell (Redox.me MM PEC 15) supplied with an illumination window as reported in the Supporting Information Figure S9. The electrolyte was a 0.5 M $Na_2SO_4$ solution (Milli-Q DI water) and the counter and reference electrodes were a 0.6 mm diameter and 250 mm length Pt wire and an Ag/AgCl (3 M KCl) electrode, respectively. The electrolyte was purged with $N_2$ during the experiments. All the electrochemical and photoelectrochemical measurements were performed by means of a Metrohm Autolab PGSTAT302N potentiostat. The current density was standardized to the electrode geometrical area.

The electrochemical active surface area (ECSA) of the photoelectrodes was estimated following the double-layer capacitance method [48-51] by analyzing the cyclic voltammograms acquired at different scan rates in the range from 2 to 50 mV/s and a voltage interval comprised between -0.05 and 0.05 mV vs. Ag/AgCl (3 M KCl) reference electrode.

For the photoelectrochemical characterization studies, photo-anode samples were illuminated from the back side of the substrate. An Oriel Instruments 66921 arc lamp and a Mightech PLS-6500 LED lamp were used as sun simulator and 6500K light sources, respectively. The power flux of both light sources was adjusted at 100mW/cm$^2$ by means of a Newport 819C-UV-2-CAL integrating sphere and a 1830R optical power meter. The 6500K LED source (410-760 nm) provides a remarkable photon flux in a defined range between 420 nm and 475 nm where the absorption edges of $BiVO_4$ and $WO_3$ semiconductors present a net absorption window (see in Figure S4 the lamp spectral distribution in relation with the diffuse reflectance spectra of the analyzed electrodes).

The choice of this 6500K light emitting LED for these experiments was made in order to suppress any significant contribution to photocurrent due to the direct excitation of the $WO_3$ semiconductor by visible and UV photons and, as seen in Figure S4, selectively excite the $BiVO_4$ semiconductor.

Experiments to ascertain the working characteristics of the $WO_3/BiVO_4$ heterojunction as a function of the $BiVO_4$ thickness were carried out by recording linear sweep voltammograms (LSV) (i.e. i-V curves between 0.0 and 1.0 V recorded at a rate of 50 mV/s) and chronoamperograms (at 1.23 V vs. RHE and for periods of time up to 5 h), under the stimulation of different light sources and illumination conditions (dark, light or chopped illumination). Tafel plots for NT and TF electrodes, the former with and without CoPi catalysts, were determined according to the Butler-Volmer equation applied to the recorded LSV diagrams. [52] Electrochemical impedance spectroscopy (EIS) analysis in the dark and under illumination was carried out to estimate the characteristics of the charge transfer mechanisms during the OER. Nysquit plot diagrams were obtained at 1.23 V vs. RHE, applying a sinusoidal signal 1 mV amplitude at frequencies from $10^5$ Hz to 0.1 Hz. Fitting analysis (Metrohm Autolab NOVA 2.1.4) of these diagrams were done assuming an equivalent circuit diagram similar to that proposed previously for this type of heterojunction electrodes. [24-26]

IPEC response of the multishell NT electrodes was determined at 0.5 V vs. Ag/AgCl (Sat. KCl) using Zahner CIMPS-QE/IPCE system consisting of a tuneable LED light source (TLS03). The IPCE measurements were performed in a three-electrode cell setup with a platinum foil as counter electrode and Ag/AgCl (KCl sat.) as reference electrode. 0.5 M $Na_2SO_4$ solution was employed as electrolyte.The IPCE values were calculated according to the following equation:

$$IPCE = \frac{I(\lambda)}{P \cdot A \cdot \lambda} \cdot \frac{h \cdot c}{q_e} \qquad \text{(Eq. 4)}$$

where $I(\lambda)$ is the photocurrent in A, $P$ is the incident light power density in W/m$^2$ at each wavelength $\lambda$ in nm, $A$ is the illuminated area in m$^2$, $h$ is the Planck constant (6.62607004 $\times$ 10$^{-34}$ J s), $c$ is the speed of light ($\sim$ 3 $\times$ 10$^8$ m/s), and $q_e$ is the elementary charge ($\sim$1.6 $\times$ 10$^{-19}$ C).

**CONCLUSIONS**

Previous results confirm the good performance of the BiVO$_4$ semiconductor as an effective sensitizing semiconductor photoanode in combination with WO$_3$. We have found that NT electrodes, prepared using vacuum deposition procedures and ONWs as substrate-supported soft-templates, present a very high electrochemically active surface area available for charge transfer reactions. The multishell ITO/WO$_3$/BiVO$_4$ structure of these electrodes provides very short and equivalent pathways along the complete length of the NTs for the photo-generated electrons to reach the ITO conductive layer. It has been also demonstrated that the thickness and homogeneity of the BiVO$_4$ outer shell layer is a second-order parameter that adjusted to a values between 65-85 nm permits to maximize the photoefficiency.

Within the range of studied photon fluxes, the WO$_3$/BiVO$_4$ heterojunction structure of the NT electrodes did not present signs of saturation, enabling the use of these electrodes under very high illumination conditions as those provided by solar collector devices. The robust character of the system (it could be reused without any noticeable loss of activity) and its long time stability make these electrodes a suitable option for practical applications, a feature that is further justified by the up-scaling possibilities of the processing methods utilized for their fabrication. Key issues by the utilized procedure are the use of vacuum processable ONWs as a scaffold and the demonstration that the MS technique can be used for the preparation of conformal layers of ITO and $WO_3$ covering completely up to 10 microns length NWs. We have also shown that the hollow multishell NT configuration is perfectly compatible with the incorporation of OER catalyst. This has been proved by the incorporation of a CoPi layer and the demonstration of an enhancement in photoelectrochemical efficiency with photocurrents of 2.23 mA cm$^{-2}$ at 1.23 V vs. RHE and one sun illumination.

**Supporting Information**.

Supporting Information section S1. Experimental Detail for the synthetic protocol.

Supporting Information, complementary Figures S1 to S9. Supporting Information, Table S1.

AUTHOR INFORMATION

**Corresponding Author**

*Corresponding author: jorge.gil@icmse.csic.es.

Jorge Gil-Rostra, Instituto de Ciencia de Materiales de Sevilla (CSIC-US). Avenida de Américo Vespucio, 49, 41092, Sevilla, SPAIN.

**Author Contributions**

The manuscript was written through the contributions of all authors. All authors have approved the final version of the manuscript.

**ACKNOWLEDGMENTS**

The authors thank the project PID2019-110430GB-C21 funded by MCIN/AEI/10.13039/501100011033 and by "ERDF (FEDER) A way of making Europe", by the "European Union". JGR also thanks the CSIC Intramural Project "Nanostructured electrodes for opto-electrochemical detection and energy applications". The project leading to this article has received funding from the EU H2020 program under the grant agreement 851929 (ERC Starting Grant 3DScavengers). Authors thank the Materials Science Institute of Seville Services for Electron Microscopy, X-Ray Diffraction and Surface Analysis, and the Center for Bioinnovation (University of Malaga) for the access to FESEM-FIB.

# SUPPORTING INFORMATION

# Photoelectrochemical water splitting with ITO/WO₃/BiVO₄/CoPi multishell nanotubes fabricated by soft-templating in vacuum

Photoelectrochemical water splitting with ITO/WO$_3$/BiVO$_4$/CoPi multishell nanotubes fabricated by soft-templating in vacuum

*J. Gil-Rostra [1]\*; J. Castillo-Seoane [1]; Q. Guo [2]; A. Jorge Sobrido [2]; A. R. González-Elipe [1;] A. Borrás [1];*

(1) Instituto de Ciencia de Materiales de Sevilla (CSIC-US). Avenida de Américo Vespucio, 49, 41092, Sevilla, SPAIN.

(2) School of Engineering and Materials Science, Queen Mary University of London, E1 4NS, London, UK

*Corresponding author: jorge.gil@icmse.csic.es

**Supporting Information section S1. Experimental Detail for the synthetic protocol.**

*(o) Substrates cleaning process:*

Commercial glass plates (25x25 mm) covered by ITO layers (supplied by Ossila, 400 nm thickness, 15 $\Omega/\square$) were used as substrates. Fused silica slides and single crystalline P-type doped (100) silicon wafers were also used as substrates for characterization purposes. All substrates were sonicated following a standard acetone (Sigma Aldrich, ACS reagent) 15 min, ethanol (Sigma Aldrich, absolute grade) 15 min, and deionized water (Milli-Q) 15 min, solvents sequence.

*(i) Radio frequency cleaning and superficial activation process:*

Once in the vacuum chamber, the substrates underwent a plasma cleaning and surface activation process. A radio frequency (RF) plasma was ignited directly on the chamber substrate holder by means of a Hunttinger PFG300 RF generator powered at 50 W (DC bias 225 V) for 15 minutes in an atmosphere of 50% $O_2$/Ar. Gases were dosed using mass flow controllers (Bronkhorst, Ar: 15 $cm^3$/min, $O_2$: 15 $cm^3$/min). The chamber working pressure was $5.0x10^{-3}$ mbar.

(ii) ITO seeds magnetron sputtering deposition process:

A very thin layer of ITO (25 nm) was deposited acting as a nucleation centers to facilitate the crystalline formation of the organic phthalocyanine nanowires. A cylindrical MS head (AJA International Inc.) equipped with a 3" ITO disc target ($In_2O_3$/$SnO_2$ 90/10, Testbourne Ltd.) was powered at 75 W using a Hunttinger PFG300 RF source in Ar atmosphere (Bronkhorst, Ar: 30 $cm^3$/min). The pressure was regulated using a butterfly valve and controlled using a Pirani pressure meter. The working pressure was kept at $5.0x10^{-3}$ mbar. The substrate holder to MS source distance was 15 cm. The substrate holder temperature was controlled at 350 °C with a spinning velocity of 20 rpm. The deposition rate and thickness was registered with a quartz crystal microbalance monitor placed close to the sample holder.

<u>(iii) Phthalocyanine (H2Pc) organic nanowires:</u>

Organic nanowires (ONWs) were growth by thermal evaporation using a LUXEL Radak thermal evaporator loaded with 29H, 31H-Phthalocyanine (Sigma Aldrich, 98%). The alumina crucible was heated from 350 ºC to 375 ºC in order to keep a uniform H2Pc flux and a constant deposition rate. The thermal evaporator temperature were operated with a PID power controller and the deposition rate were registered with a quartz crystal microbalance monitor. The deposition process was carried out in Ar (Bronkhorst, Ar: 10 $cm^3$/min). The working pressure was $1.5x10^{-3}$ mbar. The substrate holder temperature was set at 210 ºC with a spinning velocity of 20 rpm. In these conditions the H2Pc deposition rate was approximately 0.15 A/s. The evaporation process was maintained until reaching an organic precursor nominal thickness of 0.35KA.

*(iv) ITO magnetron sputtering deposition process:*

ITO was deposited by MS under operating conditions leading to a conformal layer completely covering the previously deposited ONW nanostructured templates. The deposition conditions were similar to those described in step (*ii*). In order to protect the ONW structure the substrate holder temperature was set at 210 ºC during the initial stages of the deposition and, after 10 minutes, was progressively raised up to 350 ºC. The ITO layer nominal thickness was 250 nm.

*(v) WO₃ magnetron sputtering deposition process:*

$WO_3$ films were deposited by reactive MS. A cylindrical MS head (AJA International Inc.) equipped with a 3" tungsten disc target (Testbourne Ltd., 99%) was powered at 150

W using a DC pulsed (120 kHz) Advance Energy Pinneacle + source in an atmosphere of 50% $Ar/O_2$ (Bronkhorst, Ar: 15 $cm^3$/min, O2. 15 $cm^3$/min). The working pressure was $5.0x10^{-3}$ mbar. The substrate holder temperature was set at 350 °C with a spinning velocity of 20 rpm. The process parameters (pressure, gas flow, film thickness, temperature, etc.) were controlled by the control elements previously described in sections (*i*) to (*v*). The fabricated $WO_3$ layer nominal thickness was 500 nm.

After vacuum deposition, $ITO/WO_3$ structures were annealed in air at 450 ºC (2 ºC/min) for 3 h until the complete removal of the ONWs at the core of the NTs.

*(vi) BiVO₄ drop casting deposition:*

$BiVO_4$ was incorporated by sequentially dripping on a delimited substrate area of 1.1 $cm^2$ different volumes (i.e. 20, 40, 60 μl) of a solution of $BiNO_3x5H_2O$ (Sigma Aldrich, 99.99%) (50 mM), $VO(acac)_2$ (Sigma Aldrich, 98%) (46.5 mM) and an acetic acid/ethyl acetate (9.5/0.5) mixture as solvent. Before use, the solution was agitated and periodically sonicated for 5 hours. After the complete removal of the solvent (at room temperature) the samples were annealed in air at 500 ºC (2 ºC/min) for 4 h.

*(vii) Cobalt phosphate light stimulated electro-deposition:*

The cobalt phosphate (CoPi) co-catalyst was deposited through a light stimulated electrochemical deposition method. For that, the electrodes were back illuminated with the light provided by a solar simulator (Oriel Instruments 66921) at 100 $mW/cm^2$, in a solution of 0.15 mM $Co(NO_3)_2x6H_2O$ (Sigma Aldrich, 99.99% trace metals basis) in a 0.1 M potassium phosphate buffer (at pH=7.0) applying a constant voltage of 0.4 V (vs. Ag/AgCl (3M KCl) reference electrode) for 600 s.

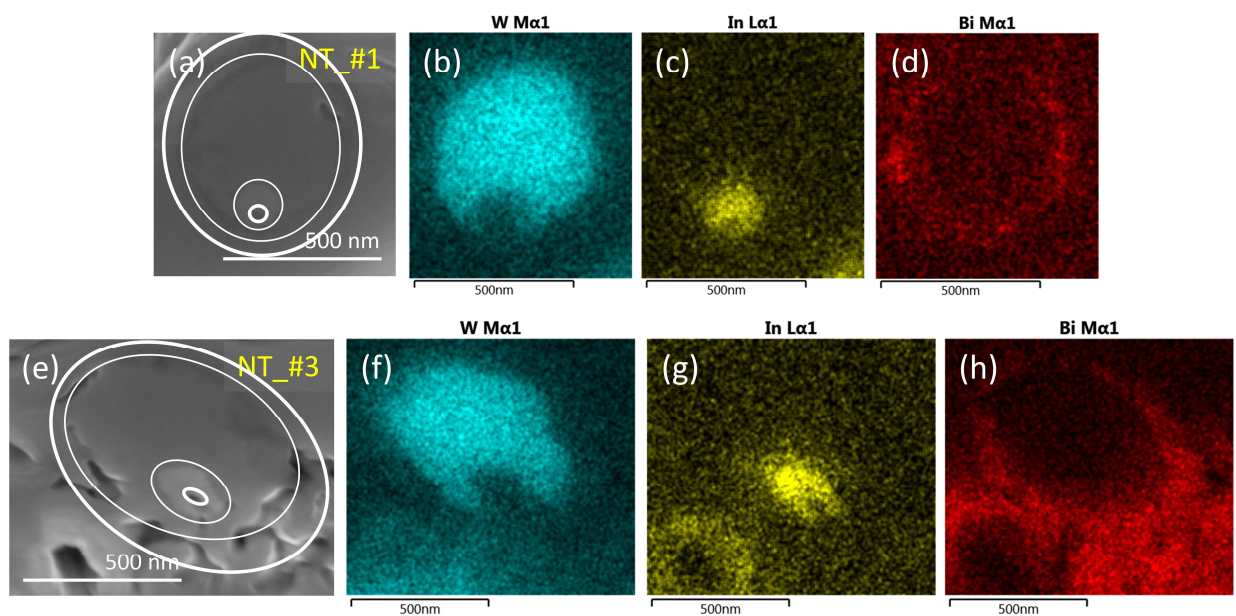

**Figure S1.** *FIB cross-section of two different nanotubes extracted from NT_#1 (a) and NT_#3 (b) and corresponding EDX mapping for representative elements W (b-f), In (c-g), and Bi (d-h).*

**Table S1.** *Thicknesses (x10), mean value and standard deviation (SD) measured from SEM micrographs for a selected set of representative nanotubes of samples NT_#0 to NT_#3.*

|        |     |     |     |     |     |     |     |     |     |     | **Mean** | **SD** |
|--------|-----|-----|-----|-----|-----|-----|-----|-----|-----|-----|----------|--------|
| **NT_#0** | 471 | 399 | 359 | 457 | 524 | 351 | 386 | 424 | 475 | 329 | **417** | 63 |
| **NT_#1** | 536 | 473 | 455 | 504 | 520 | 460 | 455 | 478 | 431 | 486 | **480** | 32 |
| **NT_#2** | 565 | 561 | 555 | 504 | 508 | 601 | 576 | 472 | 614 | 546 | **550** | 44 |
| **NT_#3** | 604 | 558 | 505 | 595 | 599 | 530 | 620 | 615 | 577 | 646 | **585** | 43 |

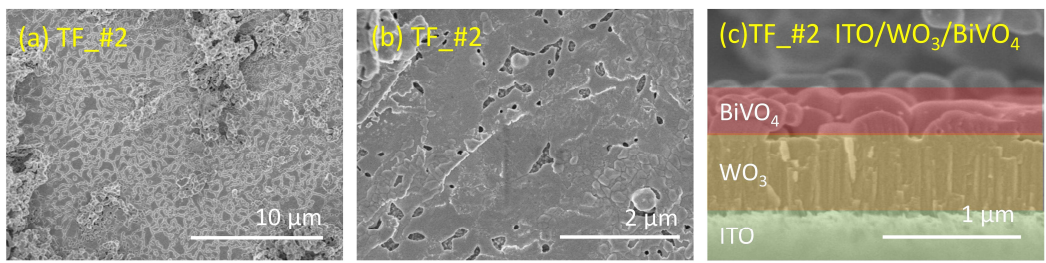

**Figure S2**. *SEM micrographs for top-views (a-b) and cross-sections (c) (colourful for easier visualization of the layers) for the thin film configuration (sample TF_#2) with equivalent BiVO$_4$ thickness to sample NT_#2.*

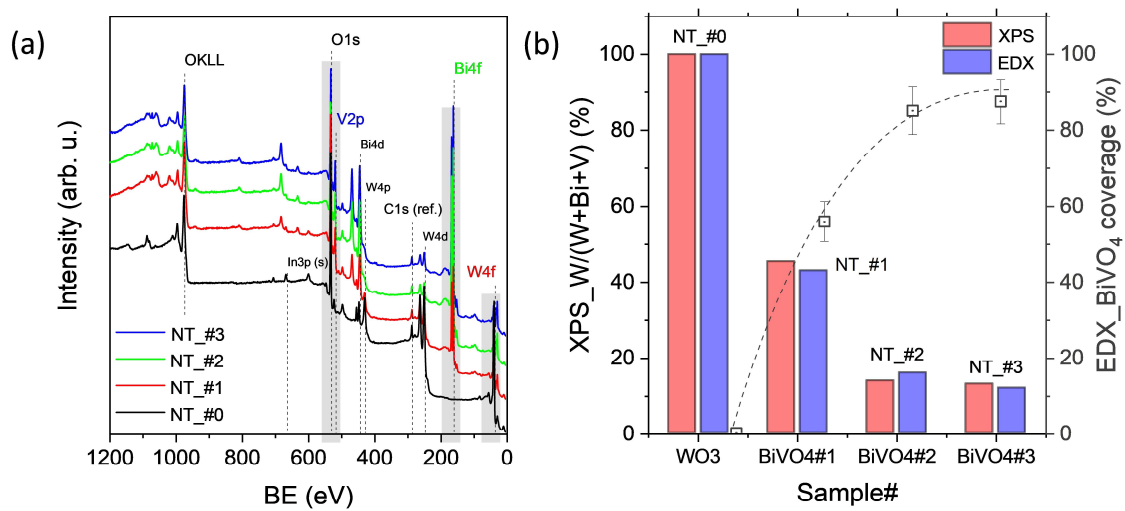

***Figure S3.** (a) XPS general spectra of samples NT_#0 to NT_#3. W4f, Bi4f and V2p photoemission signals have been highlighted. (b) Surface quantification by XPS and EDX of the W atomic ratio and the BiVO₄ degree of coverage of samples NT_#0 to NT_#3 as a function of the number of additions of the BiVO₄ precursor solution.*

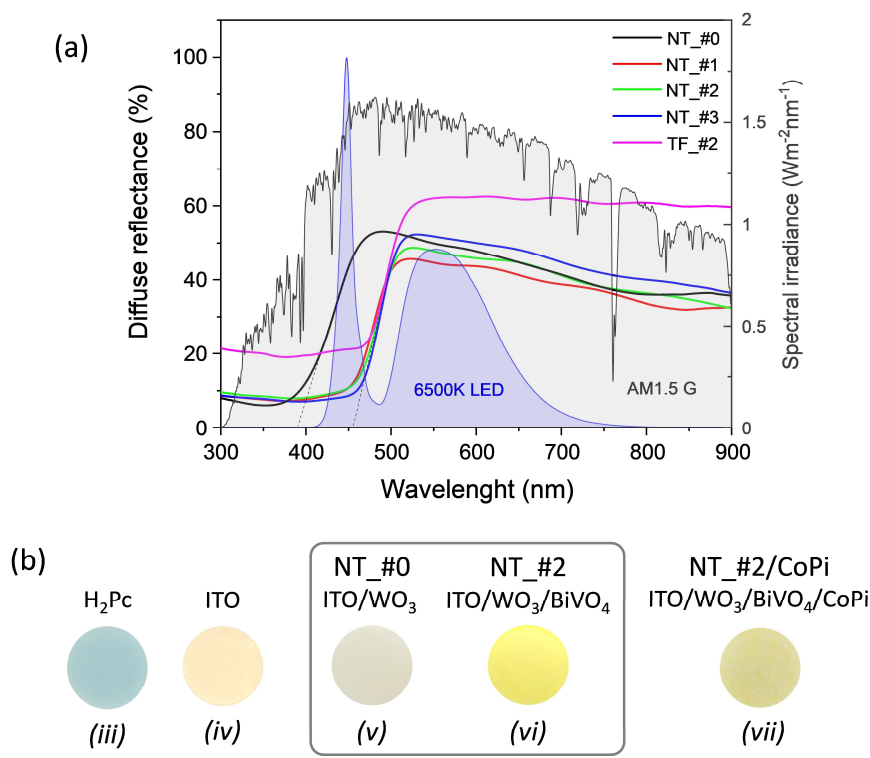

***Figure S4.*** *(a) Diffuse reflectance spectra of samples NT_#0-#3 and sample TF_#2 under illumination from the back and the front sides of the samples. Superimposed and for comparative purposes, spectral distribution of LED 6500K and AM1.5G light sources used in this work. (b) Colour evolution (actual pictures) of the in-process samples through the different steps of the photo-electrode fabrication. Phtalocyanine organic nanowires (iii), ITO nanotubes (iv), ITO/WO₃ NTs, ITO/WO₃/BiVO₄ NTs (vi), and ITO/WO₃/BiVO₄/CoPi NTs (vii).*

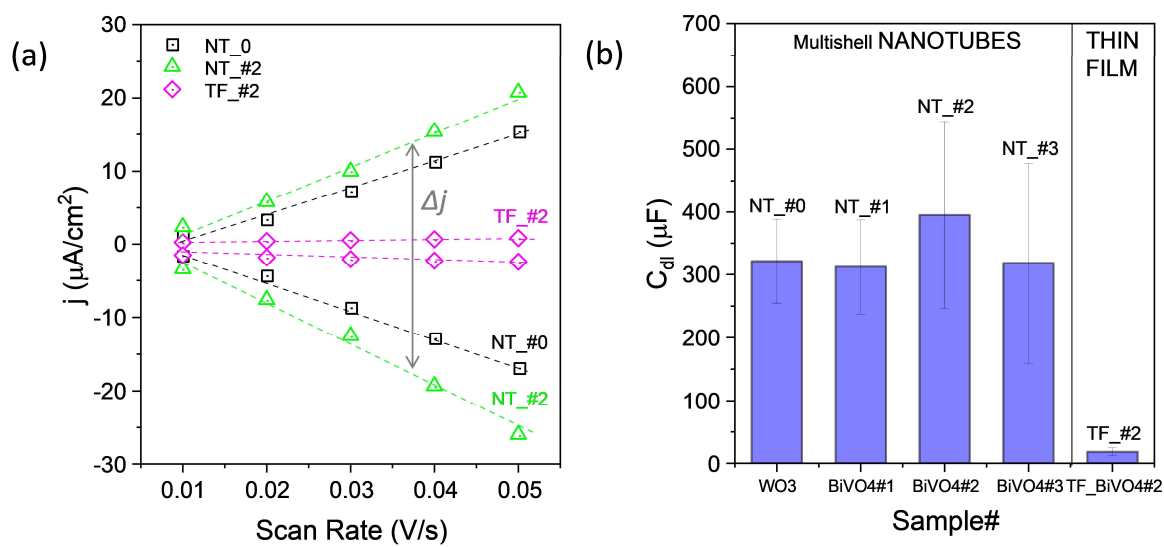

$$ECSA = \frac{C_{dl}}{C_s} \qquad C_{dl} = \frac{d(\Delta j)}{2d(SR)} \ (1)$$

***Figure S5.*** *(a) Anodic and cathodic current density values extracted from ECSA measurements (CV from -0.05 to 0.05 mV vs. Ag/AgCl (3 M KCl)) vs. scan rate (SR) for a selected set of samples NT_#0, NT_#2, and TF_#2. (b) Column diagram of the calculated double layer capacitance ($C_{dl}$) corresponding to samples NT_#0-#3 and TF_#2.*

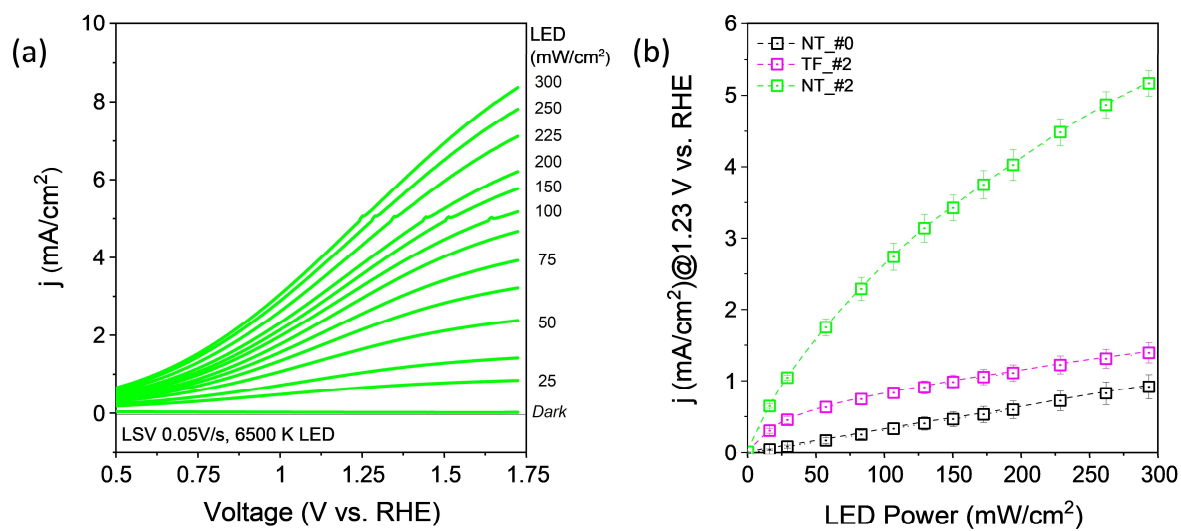

***Figure S6.*** *(a) Photoelectrochemical LSV curves measured for sample NT_#2 recorded under different power intensities of the LED 6500K illuminating lamp. (b) Photocurrent values determined for samples NT_#0, NT_#2 and TF_#2 as a function of light power intensity (LED 6500K) for a voltage of 1.23 V vs. RHE.*

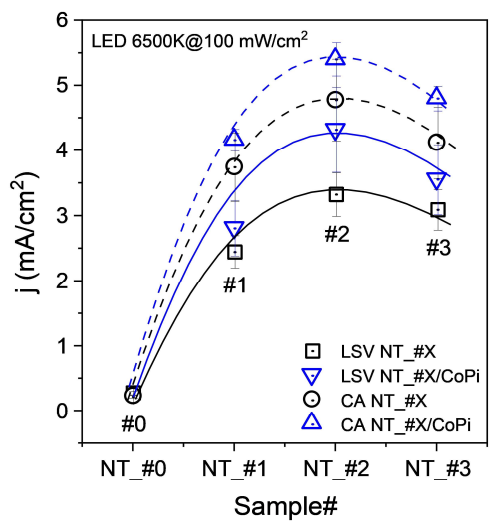

**Figure S7.** *Comparative photocurrent values measured at 1.23 V vs. RHE under 100 mW/cm² LED 6500K illumination power from LSV and CA (mean value) photoelectrochemical measurements corresponding to NT_#0-#3 and NT_#1/CoPi-#3/CoPi sample series.*

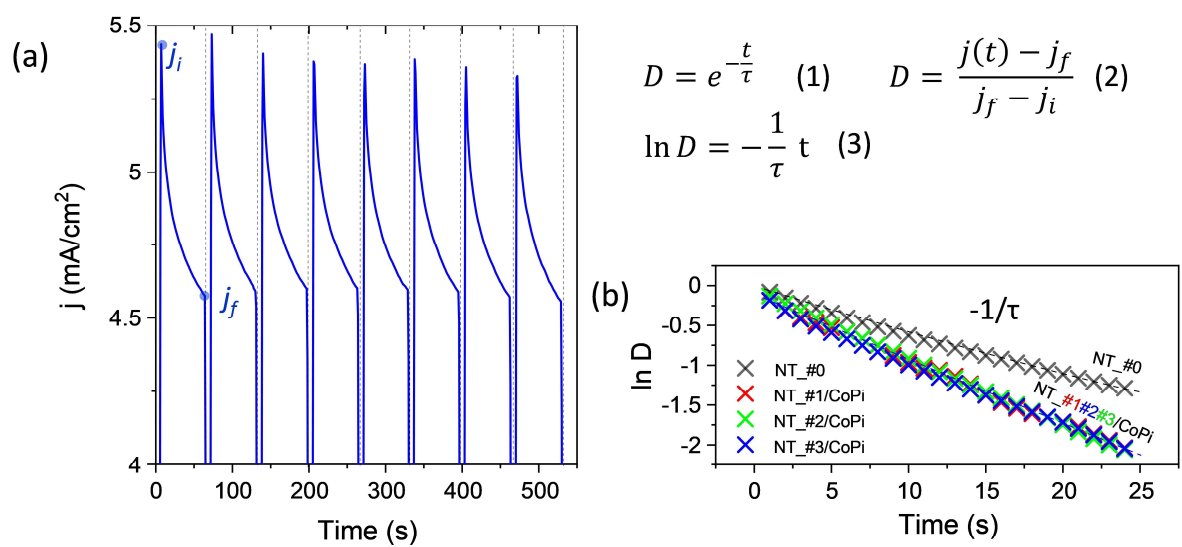

$$D = e^{-\frac{t}{\tau}} \quad (1) \qquad D = \frac{j(t) - j_f}{j_f - j_i} \quad (2)$$

$$\ln D = -\frac{1}{\tau} \, t \quad (3)$$

***Figure S8.*** *(a) Photoelectrochemical CA response of sample NT_#2 measured under chopped illumination (LED 6500K light source at 100 mW/cm², ) at 1.23 V vs. RHE. The reference parameters $j_i$ and $j_f$, defined in equation (2) as the current density at the beginning and at the end of the transient, are indicated. (b) ln D vs. t representation for samples NT_#0 and NT_#1/CoPi-#3/CoPi (b) As indicated in equation (3), the transient diffusion time (τ) can be estimated from the slope of the linear fitting of the data.*

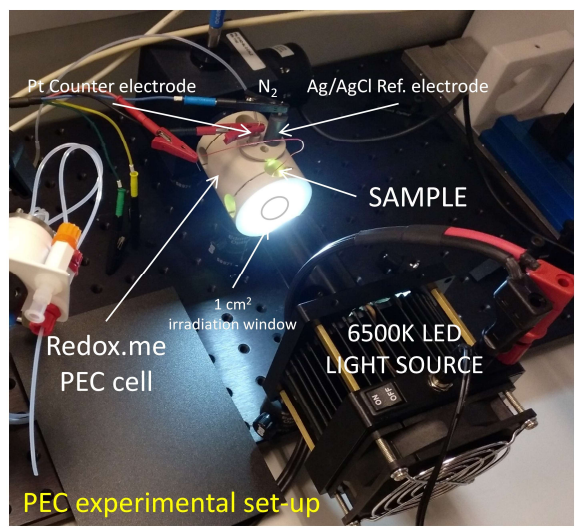

**Figure S9.** *Experimental set up used in this work for the photoelectrochemical characterization of multishell nanotubes. The system consisted of three electrodes photoelectrochemical cell (Redox.me MM PEC 15) and a 6500K LED light source (Mightech PLS-6500) or a solar radiation simulator (Oriel Instruments 66921 arc lamp, not shown here).*